\documentclass[12pt]{iopart}

\usepackage[squaren,Gray]{SIunits}
\usepackage{graphicx}
\usepackage{subfigure}
\begin{document}

\title[Mid-IR SGM design: tolerance and influence of technological constraints]{Mid-infrared sub-wavelength grating mirror design: tolerance and influence of technological constraints}

\author{C Chevallier$^{1,2}$, N Fressengeas$^2$, F Genty$^{1,2}$ and J Jacquet$^{1,2}$}
%\author{C Chevallier$^{1,2}$, N Fressengeas$^2$, F Genty$^{1,2}$ \and J Jacquet$^{1,2}$}

\address{$^{1}$ Sup\'elec, 2 Rue Edouard Belin, 57070 Metz, France}
\address{$^{2}$ LMOPS, Laboratoire Mat\'eriaux Optiques Photonique et Syst\`emes, EA 4423,\\ Unit\'e de Recherche Commune \`a l'Universit\'e Paul Verlaine - Metz et Sup\'elec,\\ 2 Rue Edouard Belin, 57070 Metz, France }
\ead{christyves.chevallier@supelec.fr}

\begin{abstract}High polarization selective Si/SiO$_2$ mid-infrared sub-wavelength grating mirrors with large bandwidth adapted to VCSEL integration are compared. These mirrors have been automatically designed for operation at~$\lambda =$~\unit{2.3}{\micro\meter} by an optimization algorithm which maximizes a specially defined quality factor. Several technological constraints in relation with the grating manufacturing process have been imposed within the optimization algorithm and their impact on the optical properties of the mirror have been evaluated. Furthermore, through the tolerance computation of the different dimensions of the structure, the robustness with respect to fabrication errors has been tested. Finally, it appears that the increase of the optical performances of the mirror imposes a less tolerant design with severer technological constraints resulting in a more stringent control of the manufacturing process.
\end{abstract}

%Uncomment for PACS numbers title message
% Keywords required only for MST, PB, PMB, PM, JOA, JOB? 
\vspace{2pc}
\noindent{\it Keywords}: Sub-wavelength grating mirror, mirror design, tolerant design, VCSEL
% Uncomment for Submitted to journal title message
\submitto{\JO}
% Comment out if separate title page not required
\maketitle

\section{Introduction}
VCSEL emitting in the mid-infrared ($\lambda >$~\unit{2.2}{\micro\meter}) are of high interest for their use as stable and tunable sources for spectroscopic measurements. In this wavelength range, AlGaInAsSb appears as the best material system for mid-infrared VCSEL fabrication. However the wavelength emission of devices made from this material system is currently limited close to~$\lambda =$~\unit{2.6}{\micro\meter} essentially due to the very thick ($>$~\unit{11}{\micro\meter}) distributed Bragg mirror (DBR) necessary to achieve 99.5~\% reflectivity~\cite{cerutti_jcg_2009, bachmann_njp_2009} which impairs the VCSEL properties. Sub-wavelength grating mirrors (SGM) with a low index sublayer~\cite{mateus_ptl_2004} can advantageously replace this DBR by exhibiting high reflectivity with a low thickness. SGM have already been successfully used as VCSEL top mirror~\cite{huang_np_2007} providing a 100~nm large stopband with a 99.9~\% reflectivity near 850~nm. A tolerance study has also been performed on this type of mirror which is made of an AlGaAs grating above an air gap as low index sublayer~\cite{zhou_ptl_2008}. The most critical parameter from a fabrication point of view is the etching of the 100~nm wide grooves. However, it has been shown that the grating parameters are very tolerant with $\pm$~30~\% on this etch length and $\pm$~10~\% on the grating period. These high tolerance values, compared to the 1~\% tolerance of DBR, is another very good advantage for VCSEL fabrication~\cite{zhou_ptl_2008}. Moreover, such mirrors present a polarization sensitivity which should improve the quality of the laser emitted beam. In a previous work, SiO$_2$/Si-based mid-infrared sub-wavelength grating mirrors (SGM) have been modelled and optimized for VCSEL application~\cite{chevallier_apa_2010}. These first simulations have shown that such SGM mirrors, centered at~$\lambda_0 =$~\unit{2.3}{\micro\meter}, can present high enough optical properties for successful VCSEL integration if an adequate design is carefully chosen.

In a first part of this work, a study of the impact of the technological constraints on the \unit{2.3}{\micro\meter} centered mirror performances is presented through the comparison of two different designs. Then, the robustness with respect to fabrication errors is evaluated thanks to a tolerance computation of the different geometrical parameters of the design.

\section{Design}
\begin{figure}
  \center
  \includegraphics[width=6cm]{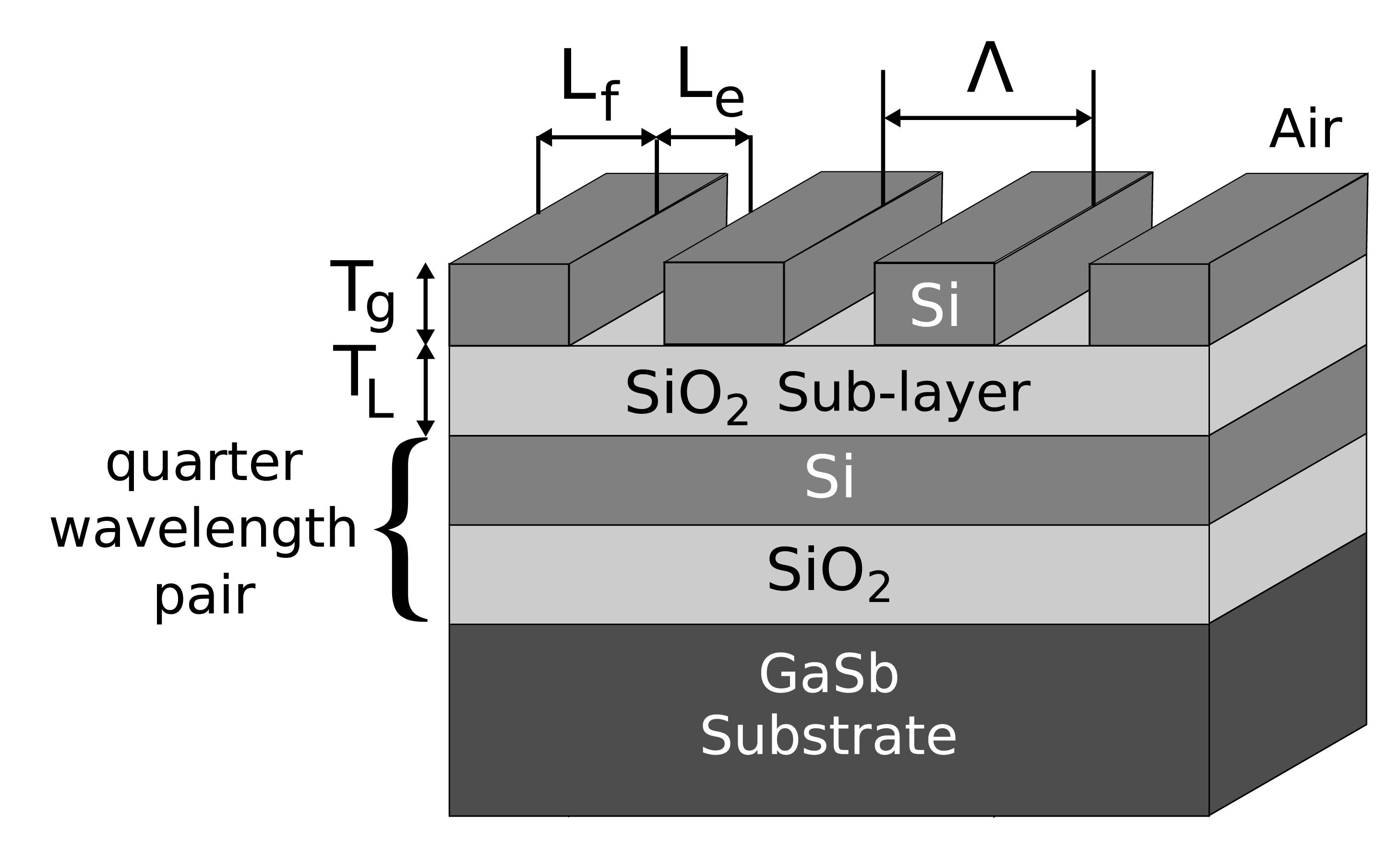}
  \caption{Scheme of the SGM combined with two quarter-wavelength layers to achieve 99.9~\% reflectivity. The structure is optimized for $\lambda =$~\unit{2.3}{\micro\meter} with optical indexes of n$_{\mbox{Si}} =$~3.48, n$_{\mbox{SiO2}} =$~1.47 and n$_{\mbox{GaSb}} =$~3.9}
  \label{figure1}
\end{figure}

The mirror structure is composed by a sub-wavelength grating mirror on top of a silica sublayer as shown on Figure~\ref{figure1}. A quarter-wavelength SiO$_2$/Si pair inserted below the grating is used to achieve the VCSEL mirror reflectivity necessary for laser operation~\cite{chevallier_apa_2010}. Then, the optical properties for a TM-polarized VCSEL have been defined. To ensure the polarization selectivity, the TM reflectivity was chosen to have a minimum value of 99.9~\% together with the widest possible bandwidth while the TE one was kept below 90~\%. These requirements allow to define a quality factor $Q$ as follow :

\begin{equation}
    Q = \frac{\Delta\lambda}{\lambda_0}\frac{1}{N}\sum_{\lambda=\lambda_1}^{\lambda_2}{R_{TM}(\lambda)g(\lambda)}
\end{equation}

The quality factor $Q$ mainly represents the normalized bandwidth of the mirror. The bandwidth $\Delta\lambda$ normalized by $\lambda_0$ is defined by the range of wavelengths $\lambda$ around $\lambda_0$ where the TM reflectivity $R_{TM}$ is higher than 99.9~\% and the TE reflectivity $R_{TE}$ is lower than 90~\%. The centering of the mirror is then taken into account by performing a gaussian weighted average of the $R_{TM}$ values on the N points of the bandwidth $\Delta\lambda~=~|\lambda_2~-~\lambda_1|$.

A global optimization algorithm~\cite{openopt} was then used to increase the SGM performances. The quality factor $Q$, specifically defined for this application, is thus automatically maximized by adjusting the different SGM characteristic dimensions. However, such a problem presents many local maxima and the use of a global algorithm is mandatory. For each evaluation of the quality factor $Q$, reflectivity spectra were numerically computed by rigorous coupled-wave analysis (RCWA)~\cite{moharam_josaa_1995,mrcwa} with constant optical index values of n$_{\mbox{Si}} =$~3.48, n$_{\mbox{SiO2}} =$~1.47 and n$_{\mbox{GaSb}} =$~3.9. Moreover, the use of an optimization algorithm allows the designer to define boundaries for the mirror geometrical dimensions. Thus, the technological constraints were also taken into account by the way of depth and length limitations. These constraints limit the filled length $L_f$, the empty length $L_e$ of the grating and its aspect ratio $AR$, defined as the grating thickness $T_g$ to the etched length $L_e$ ratio (Figure~\ref{figure1}). These limitations can be easily adjusted by the manufacturers according to the etching process used and the machine specifications.

\section{Influence of the technological constraints on the mirror performances}

SGM structures require a vertical etching profile with squared pattern to reach the 99.9~\% high reflectivity. In order to obtain an experimental square grating profile close to the theoretical one presented on Figure~\ref{figure1}, the first optimization was done by keeping the filled length $L_f$ and etched length $L_e$ of the grating larger than 500~nm and the aspect ratio $AR$ smaller than 1.1 as constraints values. This means that the optimization retains only the designs with wide grooves since a shallow pattern is easier to etch than a deep one. Compared to data already published in the literature, these constraints are relatively severe for the optimization algorithm since the best Si/SiO$_2$ SGMs with bandwidths as large as 350~nm were defined with narrower patterns of less than 260~nm and a large $AR$ of more than 2.6~\cite{mateus_ptl_2004}. Even if a simple SGM structure could exhibit 99.9~\% reflectivity under these technological constraints, the quarter wavelength layers have been added to increase the bandwidth of the mirror~\cite{chevallier_apa_2010}.% In our work, this technological choice was made in order to limit the fabrication difficulties. 

\begin{table}
\caption{\label{table1}Dimensions and tolerances of the mirror optimized under large technological constraints ($L_f$ and $L_e >$~500~nm and $AR <$~1.1).}
\begin{indented}
\lineup
\item[]\begin{tabular}{rcr@{}c@{}l}
\br
 & Optimum & \multicolumn{3}{c}{Tolerance for  $R_{TM} > 99.9$~\% and $R_{TE} < 90$~\%} \\
\mr
$L_e$ & \0675 nm & 553 nm & $< L_e <$ & \0841 nm\\
$L_f$ & \0629 nm & 539 nm & $< L_f <$ & \0685 nm\\
$\Lambda = L_e + L_f$ & 1304 nm & 1109 nm & $< \Lambda <$ & 1429 nm\\
$FF = L_f/\Lambda$ & 48.23~\% & 42.14~\% & $< FF <$ & 51.64~\%\\
$T_g$ & \0715 nm & 678 nm & $< T_g <$ & \0744 nm\\
$T_L$ & \0\017 nm & 0 nm & $< T_L <$ & \0502 nm\\
\mr
$\Delta\lambda$ & 152 nm & \\
$\Delta\lambda/\lambda_0$ & 6.6 \% & \\
\br
\end{tabular}
\end{indented}
\end{table}

The optimization of such a mirror shows a design~\cite{chevallier_apa_2010} with a grating thickness $T_g$ of 715~nm, a SiO$_2$ sublayer ($T_L$) of 17~nm, a filled length $L_f$ of 629~nm and an empty length $L_e$ of 675~nm (Table~\ref{table1}). The spectral reflection of this SGM is presented in~\cite{chevallier_apa_2010} : it exhibits a $\Delta\lambda =$~152~nm large stop-band for TM mode with a polarization selectivity ensured by a maximum reflectivity value of 80~\% for TE mode. This structure exhibits all the optical characteristics required for integration in a VCSEL structure and should limit pitfalls during the manufacturing process with large etched areas of more than 500~nm and an aspect ratio of $AR = 1.06$.

\begin{table}
\caption{\label{table2}Dimensions and tolerances of the mirror optimized under technological constraints chosen as $L_f$ and $L_e >$~400~nm and $AR <$~1.25.}
\begin{indented}
\lineup
\item[]\begin{tabular}{rlr@{}c@{}l}
\br
 & Optimum & \multicolumn{3}{c}{ Tolerance for  $R_{TM} > 99.9$~\% and $R_{TE} < 90$~\%} \\
\mr
$L_e$ & \0591 nm & 519 nm & $< L_e <$ & \0595 nm\\
$L_f$ & \0479 nm & 443 nm & $< L_f <$ & \0481 nm\\
$\Lambda = L_e + L_f$ & 1070 nm & 1015 nm & $< \Lambda <$ & 1075 nm\\
$FF = L_f/\Lambda$ & 44.77~\% & 40.07~\% &~$< FF <$~& 52.07~\%\\
$T_g$ & \0721 nm & 717 nm & $< T_g <$ & \0759 nm\\
$T_L$ & \0499 nm & 495 nm & $< T_L <$ & \0543 nm\\
\mr
$\Delta\lambda$ & 270 nm & \\
$\Delta\lambda/\lambda_0$ & 11.6 \% & \\
\br
\end{tabular}
\end{indented}
\end{table}

\begin{figure*}
  \center
  \subfigure[Reflection spectra of TM (solid blue) and TE (dashed red) mode.]{
      \includegraphics[width=7cm]{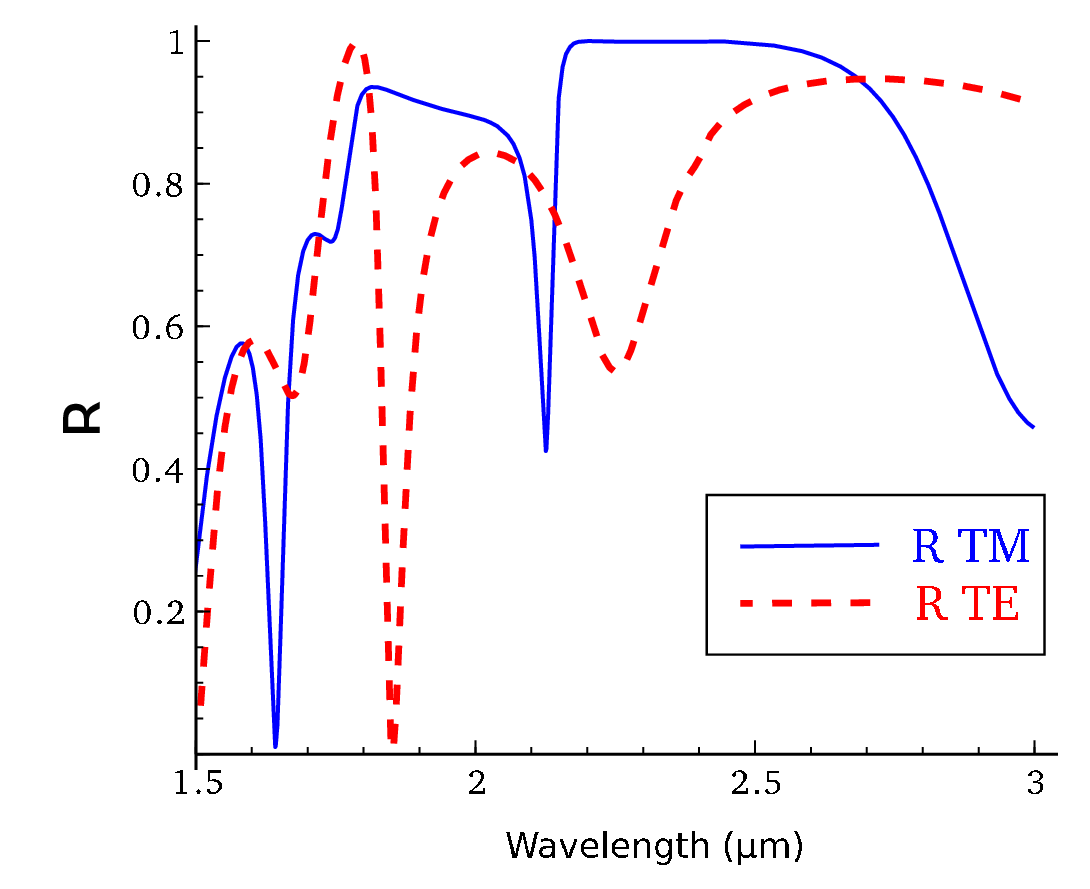}
  }
  \subfigure[Zoom on the high reflectivity values of the reflector above 99~\%.]{
      \includegraphics[width=7cm]{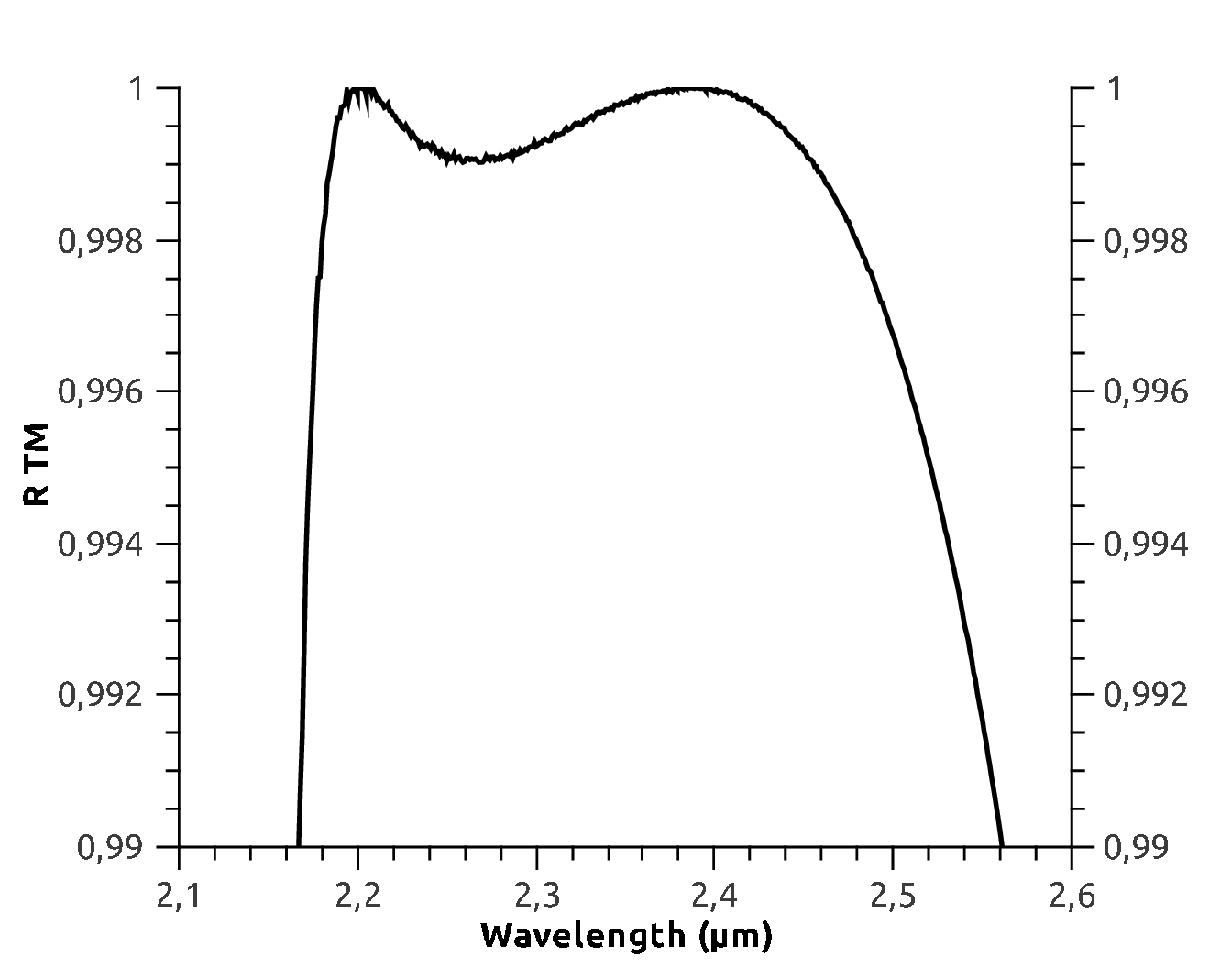}
  }

  \caption{Reflectivity of the SGM optimized under constraints ($L_e$ and $L_f >$~400~nm and $AR < $~1.25) exhibiting a 270~nm large bandwidth for a 99.9~\% TM reflectivity.}
  \label{figure2}
\end{figure*}

This mirror exhibits very interesting optical properties for VCSEL integration. However, it appears that technological constraints dramatically impact on these properties since the normalized bandwidth obtained in this case is $\Delta\lambda/\lambda_0 =$~6.6~\% while it can be 17~\% without constraints~\cite{mateus_ptl_2004}. For a better understanding and quantification of this impact, a new design was developed. For this new design, the optimization was performed under slightly more severe technological constraints which lowered the limitations of the optimization algorithm. The minimum lengths $L_e$ and $L_f$ were kept above 400~nm and the aspect ratio $AR$ smaller than 1.25. These constraints lead to a 721~nm thick grating with a filled length $L_f$ of 479~nm and an empty length $L_e$ of 591~nm. The SiO$_2$ sublayer has a thickness of 499~nm (Table~\ref{table2}). Moreover, due to these lower constraints, the reflector exhibits better performances with a $\Delta\lambda =$~270~nm bandwidth (Figure~\ref{figure2}). Nevertheless, with smaller steps for the grating and narrower and deeper grooves ($AR =$~1.22), this design requires a better control of the etching process.

To conclude this part, the quality of the mirror increases significantly with the reduction of the technological limitations where $AR = T_g/L_e$ is a key factor. However, the grating pattern could be manufactured with a certain inaccuracy resulting in new design limitations. Nevertheless, this new form of technological constraint, related to the reliability of the fabrication process, is less critical in the case of a robust design. 

\section{Robustness of the mirrors}

The robustness of a design is evaluated by performing tolerance computations on the different dimensions of the structure.

The tolerance of one parameter is defined by the variation range of this parameter for which the bandwidth of the mirror verifies the condition $\Delta\lambda >$~0~nm. This means that the design keeps $R_{TM} >$~99.9~\% and $R_{TE} <$~90~\% at $\lambda_0$. It is important to note that for the computation of one tolerance value, $L_e$ for instance, the quality of the mirror is measured by varying it while the other parameters ($L_f$, $T_g$ and $T_L$) are kept at their optimal values.

Since the different geometrical dimensions of the grating are not decorrelated, if one parameter is inaccurately achieved during the fabrication process, for instance the grating thickness $T_g$, the computed tolerances of the other parameters, in the example $L_e$, $L_f$ and $T_L$, are not valid any more. To simplify the problem in 2D, only two parameters $L_e$ and $L_f$, which variation ranges are defined by $\Delta L_e = L_e^{max} - L_e^{min}$ and $\Delta L_f = L_f^{max} - L_f^{min}$ respectively, will be considered at the following.

In the worst case, if one parameter is at its extremum value, $L_e = L_e^{max}$ for instance, a small variation of the other parameter $L_f$ would also decrease the quality of the mirror. Thus, the variation range of $L_f$ would tend to 0 when $L_e$ tends to its extremum value.

By making the hypothesis that the decrease of the variation range is linear, the four extrema $L_e^{min}$, $L_e^{max}$, $L_f^{min}$ and $L_f^{max}$ define the vertices of a quadrilateral in the ($L_e$, $L_f$)~plan. The area within the quadrilateral represents the set of pairs ($L_e$, $L_f$) which define VCSEL-quality mirrors. A similar approach can be made to define an hyper-polyhedron in dimension 4 with the parameters $L_e$, $L_f$, $T_g$ and $T_L$.

\begin{figure}
  \center
  \includegraphics[width=10cm]{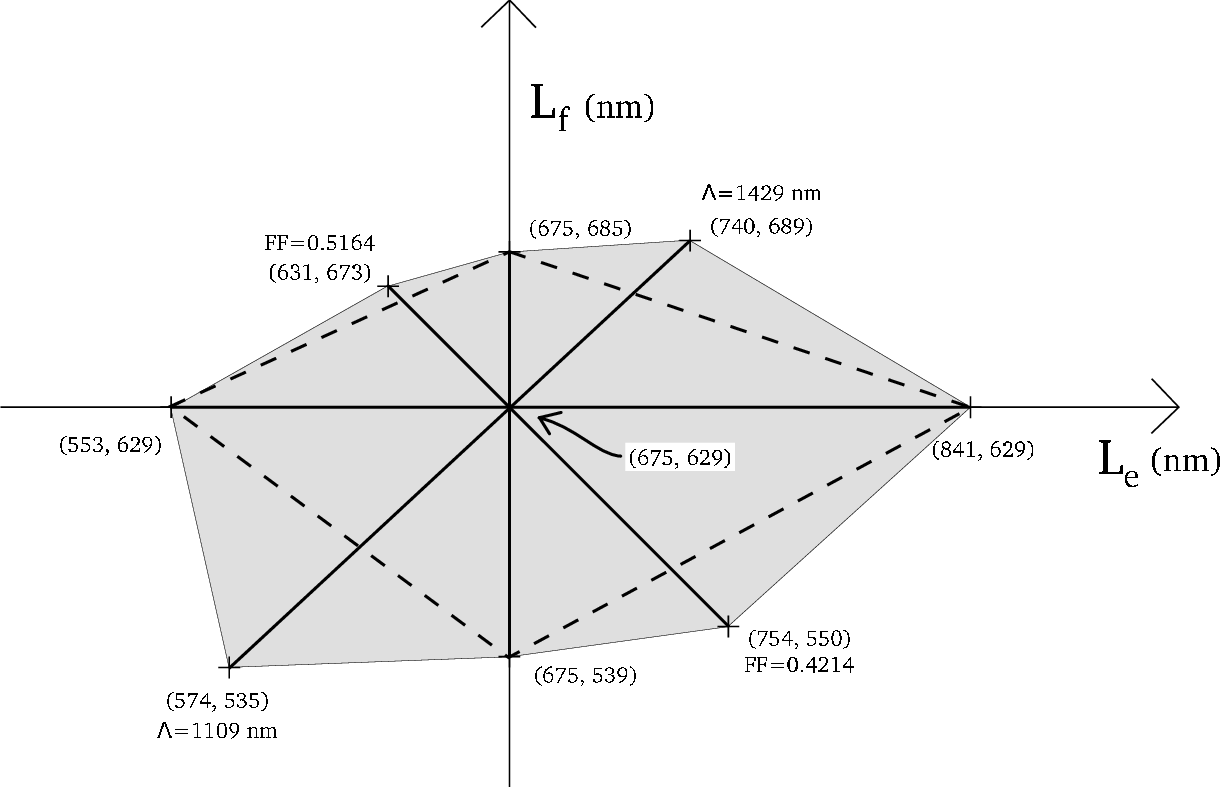}
  \caption{Tolerance of the filled length $L_f$ versus etched length $L_e$ for the first design. The point (675, 629) at the center represents the optimum value found by the optimization. The quadrilateral (dashed line) defines, in a linear approximation, the area of ($L_e$, $L_f$) pairs which correspond to VCSEL-quality designs. The computations of the period $\Lambda$ and the fill factor $FF$ define 4 points which extend the quadrilateral to a polyhedron (grey area). The worst case hypothesis is thus confirmed by a larger area of tolerance.}
  \label{figure3}
\end{figure}
\begin{figure}
  \center
  \includegraphics[width=7.5cm]{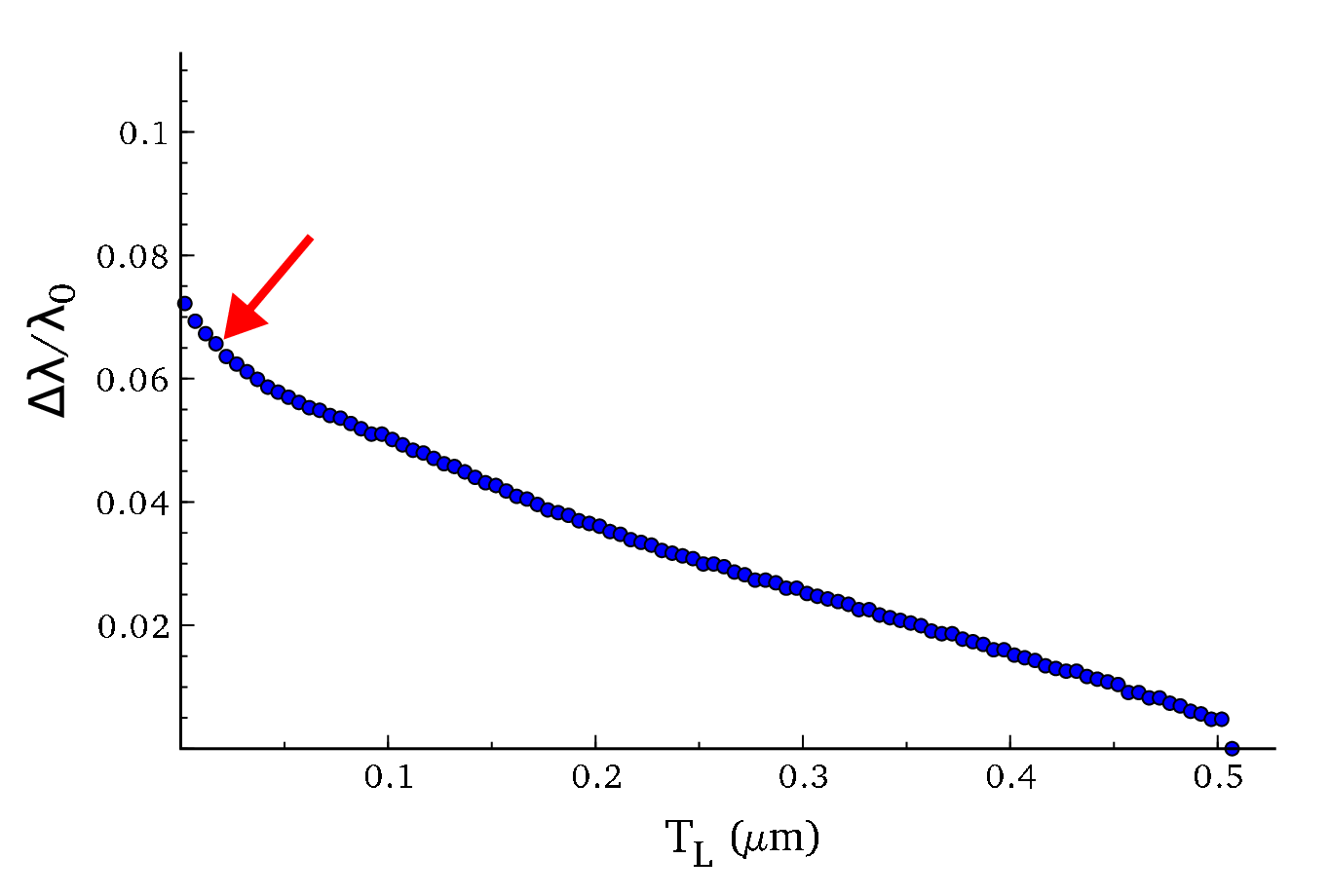}
  \caption{Evolution of the normalized bandwidth  $\Delta\lambda/\lambda_0$ versus the silica thickness $T_L$. The arrow shows the optimum design ($T_L =$~17 nm). The silica thickness $T_L$ can be as large as 100~nm without decreasing excessively the quality of the mirror by keeping a 115~nm large bandwitdh.}
  \label{figure4}
\end{figure}

The computation of the tolerances of the first design shows very large variation range of 288~nm for $L_e$, 146~nm for $L_f$ and 66~nm for $T_g$ (Table~\ref{table1}) which renders the grating very robust with respect to the fabrication imperfections. Moreover, as shown in Figure~\ref{figure3} which represents the variation range of $L_f$ versus $L_e$, the optimal value found by the optimization algorithm is well centered within the quadrilateral delimited by the extrema {(675, 685), (841, 629), (675, 539), (553, 629)} (dashed lines). The computation of the variation range of the grating period $\Lambda = L_e + L_f$ and the fill factor $FF = L_f / \Lambda$ results in ($L_e$, $L_f$) pairs which are not included within the quadrilateral. Thus, the tolerance area in the ($L_e$, $L_f$) plan is extended to a polygon, the grey area in Figure~\ref{figure3}, which is larger than the region defined by dashed line, validating the worst case hypothesis made previously.

The grating thickness is the most sensitive parameter for this design but thanks to the presence of the SiO$_2$ sublayer, a selective etching method can be used which would increase the control of the etched depth. The silica sublayer of 17~nm exhibits the largest tolerance and can vary from 0~nm up to 502~nm. As discussed in~\cite{chevallier_apa_2010}, the suppression of this layer can result in a more performant design. This is perfectly shown on Figure~\ref{figure4} where the normalized bandwidth $\Delta\lambda/\lambda_0$ is plotted versus the silica sublayer thickness $T_L$. This figure indicates that the thinner is $T_L$, the larger the bandwidth becomes. Thus the interest of this sublayer is mainly to provide the possibility of using a selective etching method and must be very thin.

\begin{figure}
  \center
  \includegraphics[width=7.5cm]{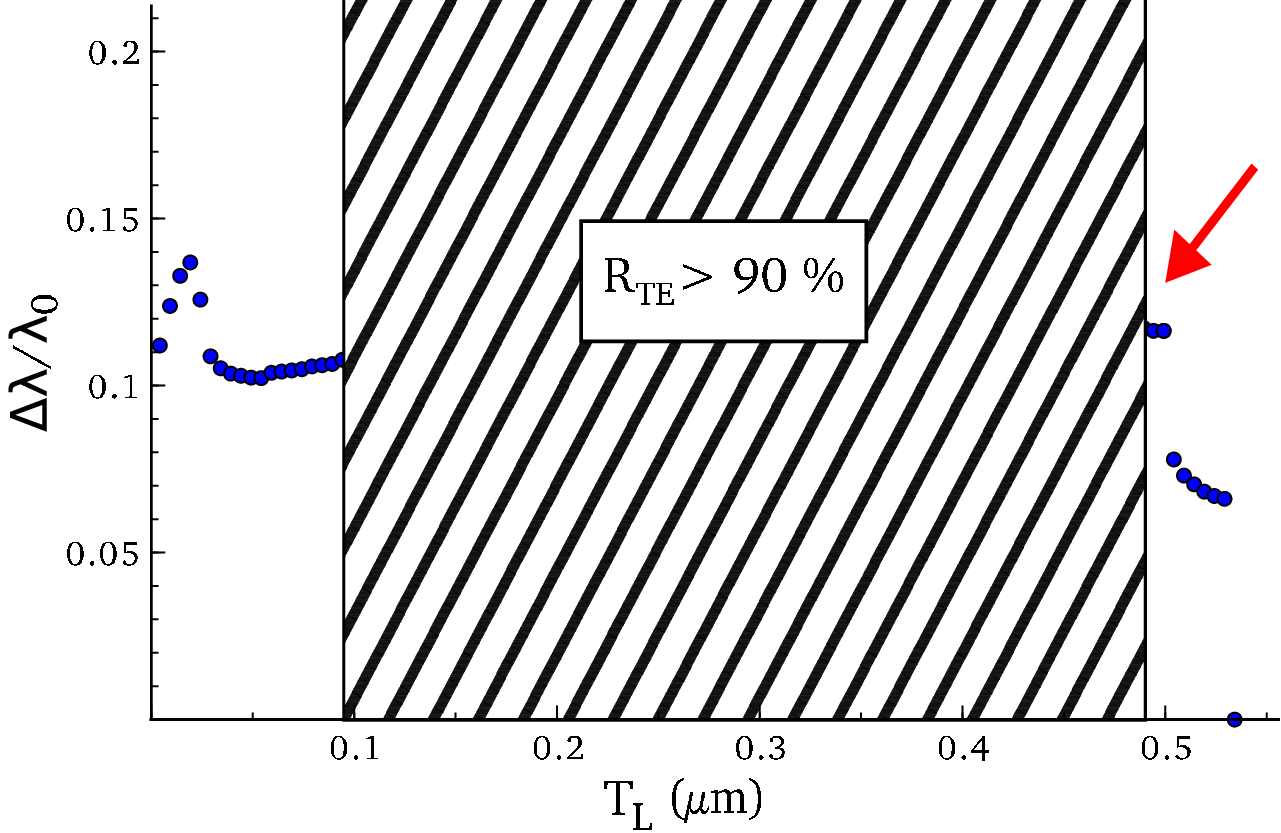}
  \caption{Evolution of the normalized bandwidth $\Delta\lambda/\lambda_0$ versus the silica thickness $T_L$ for the second design. The arrow shows the optimum design ($T_L =$~499~nm). The hatchings correspond to the silica thickness range where the TE polarization condition is not respected. For lower thicknesses ($T_L <$~100~nm), better gratings can be found but they exhibit smaller tolerance values as low as 12~nm for the filled length $L_f$.}
  \label{figure5}
\end{figure}
\begin{figure}
  \center
  \includegraphics[width=7.5cm]{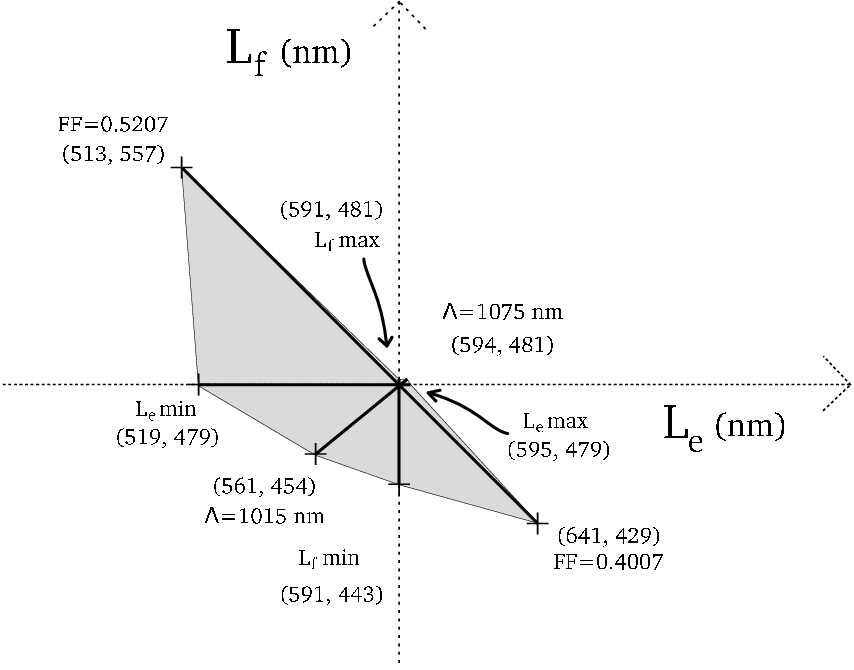}
  \caption{Tolerance of the filled length $L_f$ versus etched length $L_e$ for the second design. The optimum design (591, 479) is very closed to the boundary of the tolerance area defined by the polyhedron (grey area). This design requires a better etching precision with less than 5~nm of error allowed.}
  \label{figure6}
\end{figure}
For the second design, computations indicate smaller variation range of 76~nm for $L_e$, 38~nm for $L_f$ and 42~nm for $T_g$ (Table~\ref{table2}) which impose to have a better control of the fabrication process. Moreover, the optimum value of $T_L$ is more sensitive with a variation range of only 48~nm. This length is limited by the apparition of high TE reflectivity of more than 90~\% within the bandwidth for thicknesses $T_L <$~495~nm (Figure~\ref{figure5}) which is forbidden by our hypothesis.

Besides these smaller tolerance values, the optimum design found is not centered within the tolerance area. As shown in Figure~\ref{figure6}, the optimum point is very close to the boundary of the area in the ($L_e$, $L_f$) plan which imposes even smaller precision of less than 4~nm for the fabrication process. A solution to avoid this requirement is to choose another point in the grey area of the Figure~\ref{figure6} which would be more centered but would result in a less performant mirror.

It is also interesting to note that a smaller value of $T_L$ of 18~nm lead to a better mirror with a bandwidth as large as 300~nm (Figure~\ref{figure5}). This design has not been found by our optimization algorithm since it has smaller tolerance values as low as 12~nm for $L_f$. Such a design represents a narrow maximum for the quality function $Q$ and thanks to the use of probability in the optimization algorithm, designs with large tolerance values are statistically promoted.

\section{Conclusion}

In this work, two different SGM designs based on Si/SiO$_2$ materials and centered at \unit{2.3}{\micro\meter} are described. These mirrors with total thicknesses lower than \unit{2}{\micro\meter} are well adapted for a VCSEL integration. The design of the SGM mirror can easily be adapted to the technological constraints of the manufacturers by adjusting the limitations defined for the automated optimization process. Indeed, the comparison between different technological constraints shows that a better accuracy of the etching process results in an increase of the quality of the mirror with a 118~nm larger bandwidth. However, this quality increase is also linked to a less robust mirror with respect to fabrication imperfections on the different dimensions. The choice of the design in regard to the manufacturing process can be made by taking into account not only the limitations on the pattern resolution but also the precision required to keep a efficient mirror. This work can also be adapted to a large range of materials, such as semi-conductors, structures and wavelengths to meet the optical requirements of numerous applications.

\ack

The authors thank ANR for financial support in the framework of Marsupilami project (ANR-09-BLAN-0166-03) and IES and LAAS (France), partners of LMOPS/Sup\'elec in this project. This work was also partly funded by the InterCell grant (http://intercell.metz.supelec.fr) by INRIA and R\'egion Lorraine (CPER2007). 

\section*{References}
\bibliographystyle{unsrt}

\end{document}